# Intercomparison of the POES/MEPED Loss Cone Electron Fluxes With the CMIP6 Parametrization

H. Nesse Tyssøy[1], A. Haderlein[1], M. I. Sandanger[1], and J. Stadsnes[1]

[1]Birkeland Centre for Space Science, Department of Physics and Technology, University of Bergen (UiB), Bergen, Norway

**Abstract** Quantitative measurements of medium energy electron (MEE) precipitation (>40 keV) are a key to understand the total effect of particle precipitation on the atmosphere. The Medium Energy Proton and Electron Detector (MEPED) instrument on board the NOAA/Polar Orbiting Environmental Satellites (POES) has two sets of electron telescopes pointing ~0° and ~90° to the local vertical. Pitch angle anisotropy, which varies with particle energy, location, and geomagnetic activity, makes the 0° detector measurements a lower estimate of the flux of precipitating electrons. In the solar forcing recommended for Coupled Model Intercomparison Project (CMIP) 6 (v3.2) MEE precipitation is parameterized by Ap based on 0° detector measurements hence providing a general underestimate of the flux level. In order to assess the accuracy of the Ap model, we compare the modeled electron fluxes with estimates of the loss cone fluxes using both detectors in combination with electron pitch angle distributions from theory of wave-particle interactions. The Ap model falls short in respect to reproducing the flux level and variability associated with strong geomagnetic storms (Ap > 40) as well as the duration of corotating interaction region storms causing a systematic bias within a solar cycle. As the Ap-parameterized fluxes reach a plateau for Ap > 40, the model's ability to reflect the flux level of previous solar cycles associated with generally higher Ap values is questioned. The objective of this comparison is to understand the potential uncertainty in the energetic particle precipitation applying the CMIP6 particle energy input in order to assess its subsequent impact on the atmosphere.

## 1. Introduction

Precipitating energetic protons and electrons, ionizing the polar thermosphere and mesosphere, have long been known to initiate a series of chemical reactions increasing the production of $NO_x$ and $HO_x$ gasses (Porter et al., 1976; Rusch et al., 1981; Solomon et al., 1981). $HO_x$ and $NO_x$ gasses will destroy ozone in catalytic reactions (Andersson et al., 2014; Sinnhuber et al., 2012), and it is speculated that the subsequent change in temperature might alter stratospheric winds and wave propagation. Model simulations and meteorological reanalysis studies suggest that the energetic particle precipitation (EPP)-induced chemical-dynamical coupling could impact regional surface level climate at high latitudes during winter (Baumgaertner et al., 2011; Maliniemi et al., 2013; Rozanov et al., 2005, 2012; Seppälä et al., 2009, 2013).

In order to strengthen the understanding of the stratospheric ozone variability and its potential link to the surface climate, one has to improve on quantifying the EPP energy input at the different altitude levels of the atmosphere. Auroral electrons (<40 keV) deposit their energy in the lower thermosphere, locally increasing the production of $NO_x$ gases. The associated long lifetime implies that $NO_x$ will be transported both horizontally and vertically by background wind and waves, causing an indirect source of $NO_x$ deeper into the atmosphere in the winter polar vortex. Medium energy to relativistic electrons (>40 keV) and solar proton events (SPEs) have a direct impact on the composition of the upper stratosphere throughout the mesosphere. While the rare and sporadic SPEs have been extensively studied and fairly well quantified (Nesse Tyssøy et al., 2013; Nesse Tyssøy & Stadsnes, 2015), the fluxes of precipitating medium energy electrons (MEE) is still an outstanding question and a key to resolve the total EPP impact on the atmosphere.

Accurate quantification of the MEE precipitation, however, remains difficult due to instrumental challenges. Most of the current particle detectors in space are unsuitable for determining the amount of particles precipitating into the atmosphere (Nesse Tyssøy et al., 2016; Rodger et al., 2013). As one of the few detectors, the Medium Energy Proton and Electron Detector (MEPED) on board the Polar Orbiting Environmental Satellites (POES) and the European Organization for the Exploitation of Meteorological Satellites (EUMETSAT) MetOp, looks at pitch angles within and close to the atmospheric loss cone (LC) at the satellite orbit (~ altitude 800–850 km).

**Table 1**
*Summary of Optimized Electron Energy Channels (Ødegaard et al., 2017) and Their Sensitivity to Proton Fluxes (Evans & Greer, 2000; Yando et al., 2011)*

| Energy channel | Electron energy range | Contaminating proton energy range |
|---|---|---|
| E1 | >43 keV | 210–2,600 keV |
| E2 | >114 keV | 280–2,600 keV |
| E3 | >292 keV | 440–2,600 keV |
| P6 | >756 keV | |

MEPED consists of two electron and two proton telescopes, pointed in two directions, approximately 0° and 90° to the local vertical. At middle and high latitudes the 0° telescope measures particle fluxes that will be lost to the atmosphere, whereas the 90° telescope might detect precipitating particle fluxes and/or trapped particles in the radiation belts (Rodger, Clilverd, et al., 2010; Rodger, Carson, et al., 2010). The level of pitch angle anisotropy varies significantly with particle energy, location, and geomagnetic activity. This implies that the 0° and 90° telescopes alone cannot be used to determine the level of precipitating particle fluxes. Only in a rare case of strong pitch angle diffusion and an isotropic distribution will the 0° or 90° telescope give a realistic precipitating flux estimate. In case of an anisotropic pitch angle distribution, the 0° detector will underestimate, while the 90° detector will overestimate the flux of precipitating electrons (Nesse Tyssøy et al., 2016).

Still, with few exceptions, the MEPED 0° detector is used as quantitative measurement of the MEE precipitation (e.g., Arsenovic et al., 2016; Codrescu & Fuller-Rowell, 1997; Smith-Johnsen et al., 2017). These studies provide a useful first approximation for the minimum impact of MEE upon the atmosphere. It is, however, not always well communicated that the applied fluxes are lower estimates and that the extent of the underestimation is unclear. Van de Kamp et al. (2016) provide a model for 30–1,000 keV energetic electron precipitation (EEP) based on the 0° detector in the period 2002–2012 and empirically described plasmasphere structure. The model is both scaled to daily resolution of the geomagnetic index Ap or Dst. As a rough estimate, they point out that observed fluxes below $10^4$ to $10^5$ electrons·cm$^{-2}$·sr$^{-1}$·s$^{-1}$ may be underestimating the LC fluxes by up to a factor of about 10. The model parameterized by the Ap index has been recommended as part of the solar forcing for Coupled Model Intercomparison Project (CMIP) 6 (v3.2; Matthes et al., 2017) and as such will have a potential wide impact in studies estimating the effects of EPP upon the atmosphere. For example, it is applied in a recent study by Andersson et al. (2018), where the MEE is estimated to enhance the stratospheric response by a factor of 2. The processes initiated by the EPP are nonlinear in their nature. It is hence of great importance to understand the potential uncertainties of the impact associated with the initial sources.

In order to assess the accuracy of the van de Kamp et al. (2016) Ap model, we compare the modeled electron fluxes with the LC fluxes estimated by Nesse Tyssøy et al. (2016). By combining the electron fluxes measured by both the 0° and 90° MEPED telescopes with electron pitch angle distributions from theory of wave-particle interactions in the magnetosphere (Kennel & Petschek, 1966; Theodoridis & Paolini, 1967), a complete bounce LC flux is constructed for each of the electron energy channels >43 keV, >114 keV, and >292 keV (see Table 1; Ødegaard et al., 2017). A correction method to remove proton contamination in the electron counts is applied (Nesse Tyssøy et al., 2016; Ødegaard et al., 2016; Sandanger et al., 2015). Further relativistic electrons (>756 keV) detected by the proton detector is utilized as an extra electron energy channel (Nesse Tyssøy et al., 2016; Ødegaard et al., 2017). The following analysis will examine how well the Ap model reproduces the overall flux strength and its temporal variability at different latitudes/L shells throughout a solar cycle. Based on these findings, we discuss the limitations of using a relatively weak period of the solar cycle, the year 2002–2012, as a base for far stronger solar activity in the preceding solar cycles. The main purpose of this study is to understand the potential uncertainty in the EPP impact applying the CMIP6 particle energy input in order to assess the subsequent impact on the atmosphere.

## 2. Data and Methods
### 2.1. Constructing the LC Fluxes From POES/MEPED

MEPED is part of the Space Environment Monitor 2 (SEM-2) instrument package on board the POES and MetOp satellites, which are polar orbiting Sun-synchronous satellites at an altitude of ~850 km with an orbital period of ~100 min. The field of view of both the 0° and 90° telescopes is 30° full width.

Data handling of POES/MEPED measurement has a number of known caveats. The energy resolution is nominally >30, >100, and >300 keV in electron channels E1–E3. The detector efficiency, however, depends on the incoming energy spectrum. Ødegaard et al. (2017) determine an optimized effective integral energy limit and associated geometric factors assuming both power law and exponential spectra to give a reasonable representation of the incoming electron energies. Hence, the optimized energy limits are >43, >114, and >292 keV as listed in Table 1.

Another challenge is related to contaminating protons in the electron measurements. That means that low-energy protons hitting the detector will be counted as electrons. The nominal contaminating energy ranges (Evans & Greer, 2000; Yando et al., 2011) are listed in Table 1. Since these energies are covered by the proton telescopes, it is possible to calculate and subtract contaminating proton fluxes from the electron channels, E1–E3, as described in Nesse Tyssøy et al. (2016).

It has, however, been well documented that the solid state detectors will degrade over time as a result of radiation damage (Asikainen et al., 2012; Asikainen & Mursula, 2011; Galand & Evans, 2000; Sandanger et al., 2015). This impact is significant after 2–3 years of operation, changing the energy ranges of the proton detector. The degradation needs to be taken into account in a quantitative assessment of the data (Ødegaard et al., 2016; Sandanger et al., 2015). After correction, the electron flux measurement must still fulfill the requirement $J(E1) > J(E2) > J(E3)$. Further, in the absence of protons in the P5 channel, the presence of a relativistic electron count in rate the proton channel P6 is registered as >756 keV electron fluxes (Nesse Tyssøy et al., 2016; Ødegaard et al., 2017).

With the corrected and extended electron spectra, we can now determine the level of electron pitch angle anisotropy and diffusion using measurements from both the 0° and 90° telescopes in a combination with theoretically determined pitch angle distributions. Taking into account the detector response for different pitch angle distributions, the 0° and 90° fluxes are fitted onto the solution of the Fokker-Planck equation for particles (Kennel & Petschek, 1966). Finally, we estimate the equivalent isotropic flux level over the bounce LC in order to give a more precise estimate of the energy deposition in the upper atmosphere. The size of the LC varies from ~56° to 65° over L shell 2–10. The pointing direction of the 0° and 90° telescopes vary from 0° to 40° and 58° to 125° over the same interval, respectively. This is also illustrated by Figures A2 and A3 in Rodger, Carson, et al. (2010). A detailed description of the procedure is given in Nesse Tyssøy et al. (2016). In order to account for the detector noise level, LC fluxes are discarded whenever the associated corrected 0° electron flux drops below 250 $cm^{-2} \cdot s^{-1} \cdot sr^{-1}$ (40 counts/16 s) consistent with what is used in the Ap model we are comparing the LC fluxes with. The time resolution of the data presented in this study is 32 s (16 s active measurements).

Rejected data could potentially cause a bias in the data analysis. Hence, we do two parallel processings of the data. As a first approach, we give the rejected data a not a number (NaN) flag, and calculate the median ignoring the NaNs. However, as the majority of rejected data corresponds to low-electron fluxes, this might lead to an unfortunate bias toward overestimated electron fluxes. Pulling the bias into the safer direction yielding potentially underestimated electron fluxes can be achieved by replacing rejected data points by zero. On a physical basis, however, continuing with a zero median neglects the nonzero effects of the minority high-flux data, which are entirely ignored by applying the median. Therefore, determination of daily fluxes based on the data set replacing rejected data by zero uses daily means instead. The two approaches correspond to "nan-median" (removing data points and applying the median) and "zero-mean" (replacing data points by zero and continue with applying empirical means). The consistency between the results from the two approaches, or lack thereof, gives an indication of the uncertainty in estimating the daily fluxes.

### 2.2. Medium Energy Parameterization Used in CMIP6

The new solar forcing data set for CMIP6 includes for the very first time MEE. The electron flux data are provided by a parameterization based on the Ap index. The general principle of the model is to use the 0° detector flux data acquired in the period 2002–2012 to fit equations of the integral electron flux >30 keV and the spectral power law gradient, $k$. The flux data are binned with respect to their L value including 2–10, with a resolution of 0.5, and in 3-hr UT intervals. There is no distinction between different MLT sectors. Median electron fluxes are determined for each bin and linearly averaged in order to calculate daily fluxes. This is done for each energy channel. The data points in all three channels are replaced by zero whenever the >30 keV electron flux lies below the noise level, which is defined to be 250 electrons·$cm^{-2} \cdot s^{-1} \cdot sr^{-1}$. The remaining data points are then used to fit a power law spectral function for each day and L bin. Based on the obtained spectral gradient, the >30-keV flux, F30, is calculated and referred to as the modified POES data.

Next, the Ap predictor is fitted to the obtained k and F30 data points by a simple least squares fitting routine. The fitting functions describe k and F30 entirely based on the respective daily geomagnetic index and the chosen L shell. The model equations based on the Ap index are presented in equations (8) and (9) in van de Kamp et al., 2016 and given below:

$$F_{30} = \frac{e^A}{e^{-b(S_{pp}-s)} + e^{c(S_{pp}-s)} + d},$$

with

$$A = 8.2091 \text{Ap}^{0.16255}, \quad b = 1.3754 \text{Ap}^{0.33042}, \quad c = 0.13334 \text{Ap}^{0.42616}, \quad s = 2.2833 \text{Ap}^{-0.2299},$$
$$d = 2.7563 \times 10^{-4} \text{Ap}^{2.6116}, \quad S_{pp} = L - L_{pp}, \quad L_{pp}(t) = -0.743 \ln\left(\max_{t-1,t} \text{Ap}\right) + 6.5257$$

and

$$k = \frac{-1}{Ee^{-bS_{pp}} + 0.3045 \cosh(0.20098(S_{pp}-s))} - 1$$

with

$$E = 3.3777 \text{Ap}^{-1.7038} + 0.15, b = 3.7632 \text{Ap}^{-0.16034}, s = 12.184 \text{Ap}^{-0.30111}$$

A general advantage of the Ap model is its long temporal coverage of up to 100 years. Ambiguous situations, for example, during SPEs when proton contamination of the original data is likely, are bypassed by relying on geomagnetic indices.

## 3. POES/MEPED LC Electron Fluxes and the CMIP6 MEE Parametrization

### 3.1. Overall Flux Strength

As a first step, the overall performance of the Ap model with regard to reproducing the general flux strength is tested. The left column of Figure 1 illustrates the electron flux distribution with respect to different L shells and Ap values as given by the Ap model functions (van de Kamp et al., 2016). The lower energy limits are given as >43 and >114 keV in order to match the MEPED optimized energy thresholds using the energy spectral gradient provided by the model. The modeled fluxes grow stronger with increasing Ap values reaching red colors corresponding to flux levels of >5·10$^5$ cm$^{-2}$·sr$^{-1}$·s$^{-1}$ for >43 keV and >1·10$^4$ cm$^{-2}$·sr$^{-1}$·s$^{-1}$ for >114 keV. In order to compare the MEPED LC fluxes with the model, daily fluxes were determined in accordance with the approach described in van de Kamp et al. (2016). Data were binned into 3-hr intervals and L shell bin widths of 0.5. For each bin the median flux was calculated, corresponding to zonal averaging. Subsequently, eight 3-hr median fluxes were linearly averaged in order to obtain daily flux values. The overall shape of the colored patterns seems quite similar, with the model well reproducing the region of lower fluxes for low Ap and high L in the E2 channel. However, LC fluxes surpass modeled fluxes by 1 order of magnitude in high flux regions and the ratio between them is apparently not constant for high and low fluxes. In the low flux regions, for instance, modeled fluxes are almost 2 orders of magnitude smaller than LC fluxes. Thus, both in the E1 and E2 channels, LC fluxes exceed modeled fluxes significantly, while the degree of underestimation varies with different flux levels. Again, this relates to the different degrees of pitch angle diffusion depending on geomagnetic activity, but we should also note that the LC flux routine is more uncertain for lower fluxes.

### 3.2. Yearly Variability

With Ap as the only predictor of the electron flux level, the model assumes a direct link between the level of precipitating electron fluxes at all energies and Ap. Still, EEP events are usually associated with two different solar wind drivers, coronal mass ejection (CME) associated with sunspots and corotating interaction region (CIR) associated with high speed solar wind streamers from coronal holes. Their occurrence varies with the solar cycle. Based on solar cycle 21, Richardson et al. (2000) found that CIR storms make their largest contribution to geomagnetic activity, ~70%, during the declining phase of the solar cycle, and they account for ~30% of the geomagnetic activity during solar maximum. The geomagnetic intensity of the resulting storm depends on the combination of solar wind speed and Interplanetary Magnetic Field (IMF) southward $B_z$ component (Gopalswamy, 2008). The two types of events will therefore have quantitatively different geomagnetic signatures (Tsurutani et al., 2006) as the geomagnetic activity is manifested differently in the various forms of geomagnetic activity (ring current, convection, radiation belt, aurora etc.) as well as storm duration (Borovsky & Denton, 2006).

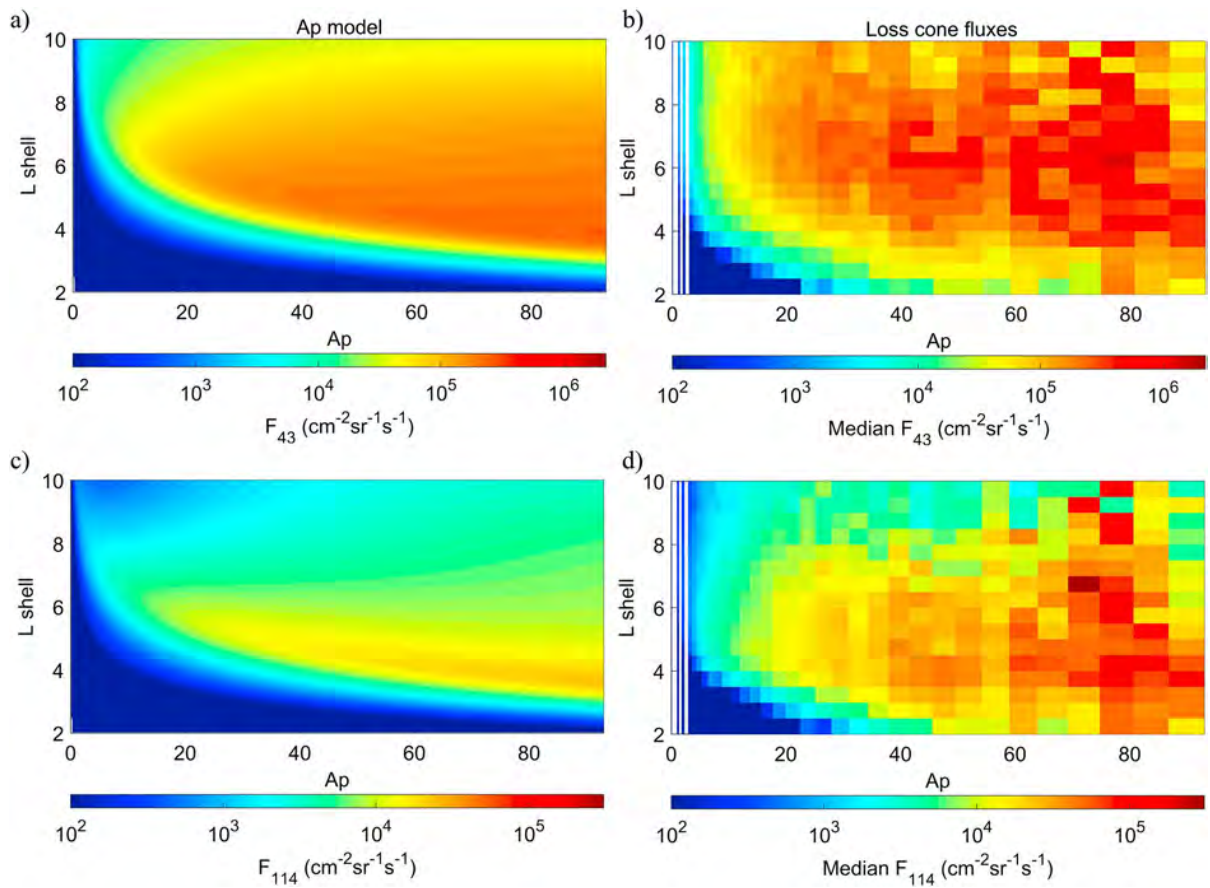

**Figure 1.** (left column) The Ap-modeled integral fluxes >43 keV (top) and >114 keV (bottom) as function of L shell and Ap. (right column) The median of the Medium Energy Proton and Electron Detector loss cone integral fluxes >43 keV (top) and >114 keV (bottom) based on all data from all magnetic local time sectors of the years 2003–2012 as function L shell and Ap.

In order to assess whether the occurrence rate of the different types of storms will cause a systematic bias using the model throughout a solar cycle, the LC fluxes during the years 2003, 2005, and 2008 are examined separately and compared. The year 2003 was located close to the solar cycle peak, whereas 2005 and 2008 lay in the declining and minimum phase, respectively.

The year 2003 is characterized by a high number of HSSWS and CIR events (Zhang et al., 2008), which despite some strong CME events dominate the annual contributions to EEP fluxes (>30 keV). The relative contribution of HSSWS and CME to EEP fluxes (>30 keV) is approximately equal in 2005, whereas 2008 is a year without significant CME activity (Asikainen & Ruopsa, 2016). Figure 2 shows POES/MEPED >43 keV LC fluxes binned according to Ap index and L shell separately for the three years. One striking feature is the strong difference in the Ap values in 2008 compared to the other two years. Whereas the Ap index reaches values of >90 in 2003 and 2005, it does not surpass a value of 40 in 2008. For Ap values <40, the general color distribution is almost undistinguishable for the different years, but slightly weaker at low L shells in 2008. There is, however, not a smooth variability with Ap for the different L shells as the fluxes may vary an order of size with a few Ap values. In particular, we note that in 2008 for L shells 4–6 flux values of >5·$10^5$ cm$^{-1}$·sr$^{-1}$·s$^{-1}$ and >1·$10^5$ cm$^{-1}$·sr$^{-1}$·s$^{-1}$ for >43 keV and >114 keV are already reached for rather low Ap values (<37), whereas in 2003 and 2005 Ap values are larger before the same flux levels are obtained. Furthermore, the dark blue area of low fluxes in low L shells stretches to significantly higher Ap values in 2008 compared to the figures referring to years 2003 and 2005. This is consistent with EEP events driven by CIR storms, which are not typically associated with large expansions of the auroral oval.

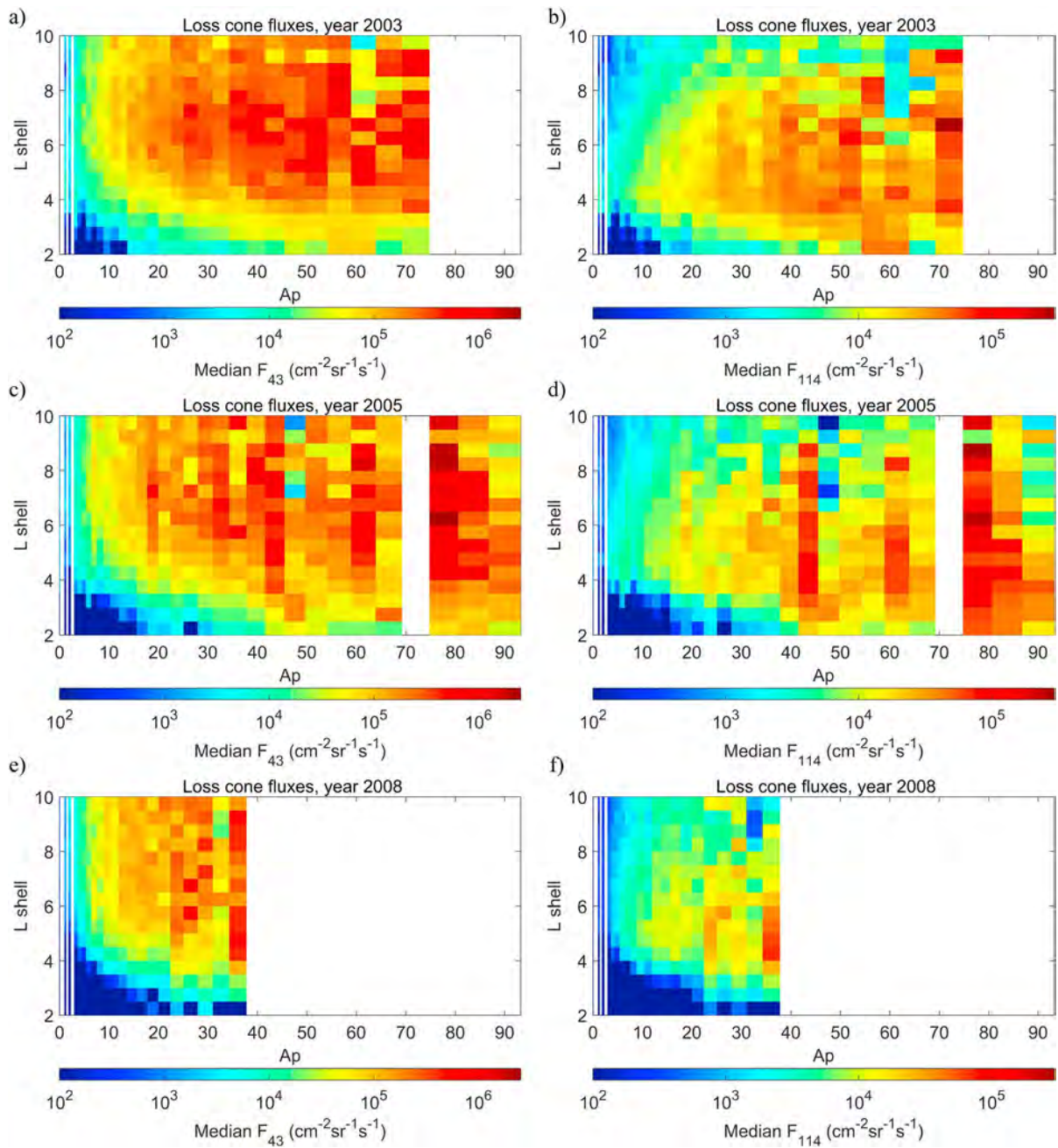

**Figure 2.** The median of the Medium Energy Proton and Electron Detector loss cone integral fluxes >43 keV (right) and >114 keV (left) based on all data from all magnetic local time sectors as function of L shell and Ap separately for the years (a. b) 2003, (c, d) 2005, and (e, f) 2008.

Table 2 shows a comparison of the nan-median (upper table) and zero mean (lower table) LC daily fluxes to the Ap model daily fluxes for >43 and >114 keV throughout the years 2003, 2005, and 2008 at L shells 5.0–5.5. The general median flux level decreases from 2003 to 2008 for all flux estimates. For >43 keV, the

**Table 2**
*Median Fluxes and Flux Ranges (Max-Min) of the (a) POES LC Nan-Median and (b) POES LC Zero-Mean and Ap Model Data Sets in Years 2003, 2005 and 2008 Shown for >43 keV and >114 keV*

| Year | POES LC flux (cm$^{-2}$·s$^{-1}$·sr$^{-1}$; daily: nan-median) | | | Ap model flux (cm$^{-2}$·s$^{-1}$·sr$^{-1}$) | | | Difference Median | Ratio Median |
|---|---|---|---|---|---|---|---|---|
| | Median | Max | Min | Median | Max | Min | | |
| **>43 keV** | | | | | | | | |
| 2003 | 87,000 | 1,700,000 | 680 | 62,000 | 290,000 | 74 | 29,000 | 0.45 |
| 2005 | 25,000 | 990,000 | 140 | 5,400 | 220,000 | 24 | 12,000 | 0.28 |
| 2008 | 9,600 | 730,000 | 100 | 1,000 | 190,000 | 24 | 6,500 | 0.17 |
| **>114 keV** | | | | | | | | |
| 2003 | 12000 | 580,000 | 300 | 6,200 | 13,000 | 50 | 6,100 | 0.35 |
| 2005 | 5300 | 96,000 | 25 | 1,020 | 13,000 | 17 | 3,200 | 0.26 |
| 2008 | 2500 | 64,000 | 10 | 338 | 13,000 | 17 | 1,800 | 0.20 |
| Year | POES LC flux (cm$^{-2}$·s$^{-1}$·sr$^{-1}$) (daily: zero-mean) | | | Ap model flux (cm$^{-2}$·s$^{-1}$·sr$^{-1}$) | | | Difference Median | Ratio Median |
| | Median | Max | Min | Median | Max | Min | | |
| **>43 keV** | | | | | | | | |
| 2003 | 140,000 | 1,650,000 | 2,400 | 62,000 | 290,000 | 74 | 73,000 | 0.32 |
| 2005 | 59,000 | 1,000,000 | 1,000 | 5,400 | 220,000 | 24 | 35,000 | 0.12 |
| 2008 | 22,000 | 800,000 | 28 | 1,000 | 190,000 | 24 | 21,000 | 0.08 |
| **>114 keV** | | | | | | | | |
| 2003 | 12,000 | 420,000 | 65 | 6,200 | 13,000 | 50 | 6,300 | 0.34 |
| 2005 | 4,400 | 160,000 | 11 | 1,020 | 13,000 | 17 | 2,600 | 0.26 |
| 2008 | 1,600 | 110,000 | 7.6 | 338 | 13,000 | 17 | 1,100 | 0.26 |

*Note.* Median absolute differences (POES model) and ratios (Ap model/POES LC) based on a day-to-day variability are shown in the last column. Data from L shells between 5 and 5.5 were considered. All MLT sectors were included.

zero-mean (nan-median) LC fluxes decrease by a factor of ~6 (~9), while the Ap model fluxes decrease by a factor of ~60. For >114 keV, the zero-mean (nan-median) LC fluxes decrease by a factor of ~7 (~5), while the Ap model fluxes decrease by a factor of ~18. The Ap have yearly mean values that drop from 22 to 7 from 2003 to 2008, respectively. Table 2 also lists the median absolute difference (MEPED LC fluxes − Ap model fluxes) and ratio (Ap model fluxes/MEPED LC fluxes). As expected due to higher general flux values, the MEPED LC data sets contain fluxes with a significantly wider range than the Ap model data set.

In summary, it appears that the Ap model exaggerates the flux level difference between solar maximum and solar minimum compared to the median and mean LC flux levels.

### 3.3. Day-To-Day Flux Variability

In order to assess the Ap model's ability to reflect day-to-day flux variations, the temporal evolution of modeled fluxes is compared to MEPED >43 keV LC fluxes on different L shells. Figure 3 contains separate plots showing fluxes in years 2003, 2005, and 2008. The three upper panels in each subfigure depict fluxes (on the $y$ axis) on L shells 4.25, 5.25, and 7.75, corresponding to approximately 61°, 64°, and 69° geomagnetic latitude, respectively. The daily Ap index of the respective year is shown in the lowest plot of each figure. All panels share a common $x$ axis stating day of year (DOY) in the three years. The black line depicts fluxes obtained from applying the Ap model to the respective daily Ap index. MEPED LC flux variations are depicted by two lines, one blue and the other red, referring to "nan-median" and "zero-mean", respectively.

Apart from the general offset in flux levels, the fluxes shown in Figure 3 seem to evolve quite coherently. Short-term variability during periods of high Ap is, however, not captured by the model as can be seen, for example, in the two middle panels in Figure 3a. The entire year 2003 is characterized by high Ap values and strong geomagnetic activity. Especially during the Halloween SPE centered around DOY 300, long periods of strongly elevated Ap are present. Nonetheless, clear Ap peaks are visible. The corresponding MEPED LC fluxes show variations both for the "nan-median" and the "zero-mean" data set. Examination of modeled fluxes in these periods shows, on the other hand, plateau-like features with little variability. The reason for this behavior lies with the nonlinearity of the Ap dependence in the model. For elevated Ap values, modeled fluxes saturate which suppresses variations that are clearly observed in the LC fluxes. Thus, the same flux levels in the 5.25 L shell are obtained for all Ap values above approximately 40. Looking specifically at the rise in Ap around DOY 285 in 2003, fluxes in this L shell reach maximum values and saturation immediately and stay there for many days despite a sudden drop of the Ap index by 50% directly after the peak. This issue is less present in 2005 as the Ap index exhibits rather single peaks followed by drastic drops than extended periods of elevated activity. Due to the low Ap level in 2008, flux saturation does not occur at all.

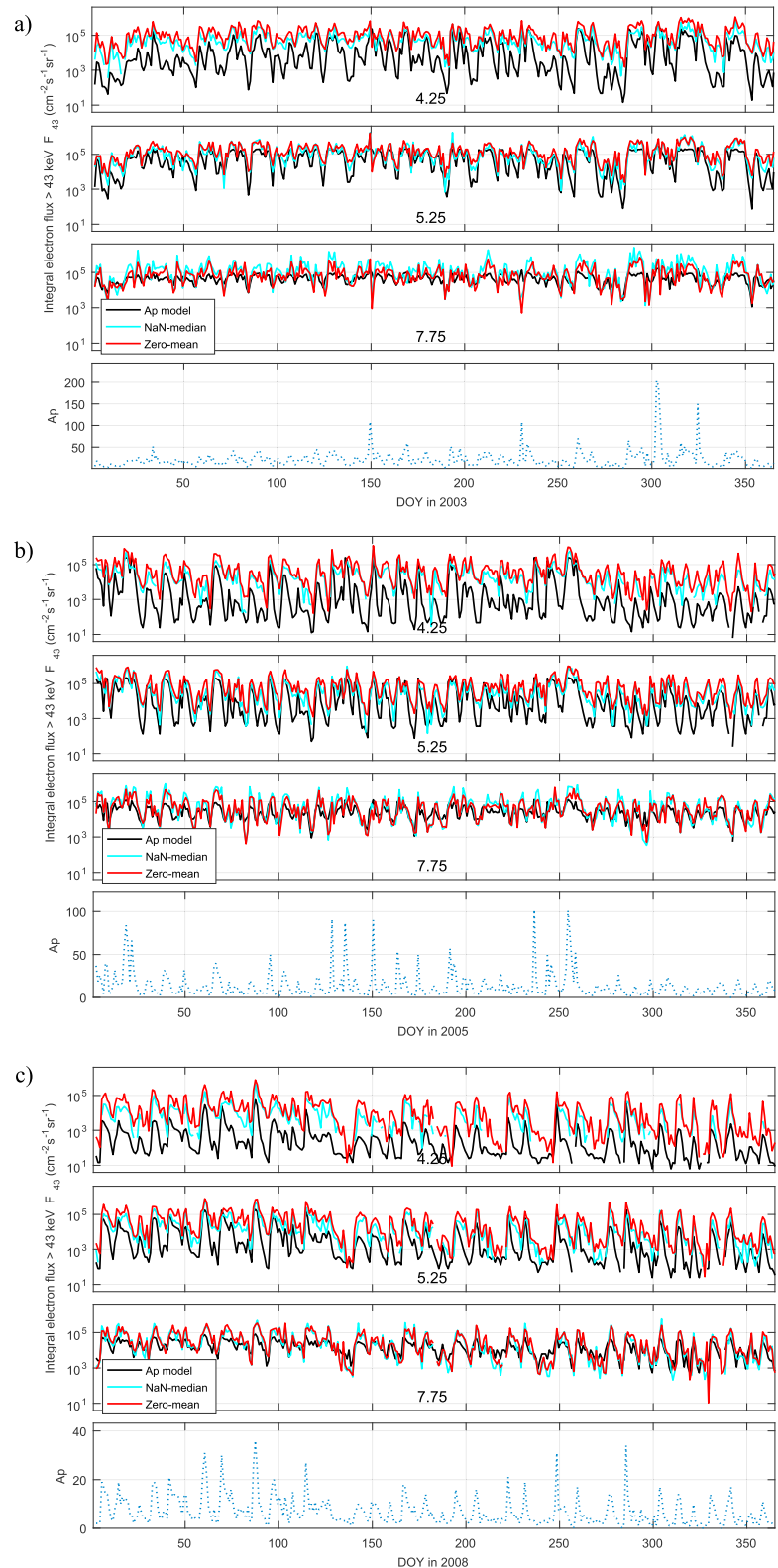

**Figure 3.** Integral fluxes >43 keV based on the Ap model and MEPED LC fluxes in years (a) 2003, (b) 2005, and (c) 2008 for three different L shell values, 4.25, 5.25, and 7.75. The MEPED LC fluxes are calculated using median treating data below the noise floor to a not a number and by calculating mean values treating data below the noise floor as zero value. The daily Ap index is shown in the lower panels. MEPED = Medium Energy Proton and Electron Detector; LC = loss cone.

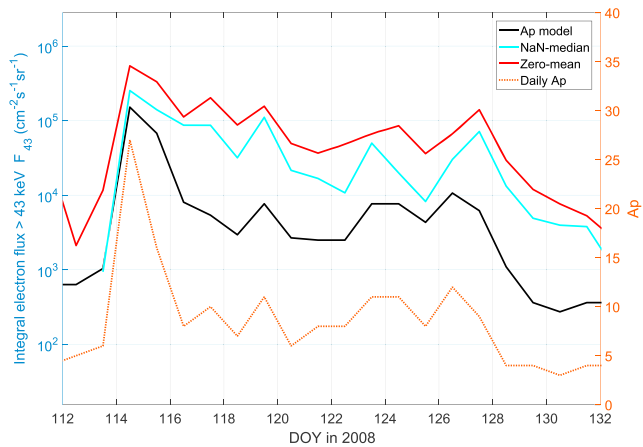

**Figure 4.** Integral fluxes >43 keV based on the Ap model, Medium Energy Proton and Electron Detector loss cone fluxes calculated using nan-median and zero-mean during a corotating interaction region storm in 2008 (DOY 112–132) on L shells 5–5.5. Daily Ap values are illustrated on the right-hand *y* axis. DOY = day of year.

On the other hand, 2008 is characterized by a series of CIR storms. In contrast to CME storms, CIR storms are characterized by significantly lower geomagnetic perturbations and lower Ap values. An example is given in Figure 4 for a typical CIR storm in 2008. The peak Ap value does not exceed 27 and the modeled Ap fluxes nicely follow the rise in Ap on DOY 114. Differences in peak flux levels between the Ap model and MEPED LC measurements persist but are expected due to the likely underestimation of LC fluxes by the 0° detector. This discrepancy, is, however, increased during the recovery phase of the storm. Bound to the Ap index, the modeled fluxes are drawn toward lower electron fluxes whereas both nan-median and zero-mean fluxes stay on higher flux levels and exhibit a slower decrease.

In order to obtain numbers that state the degree of coherence between modeled and zero-mean LC flux data, the correlation coefficient for the two data sets is calculated for different years, latitudes, and energies. The resulting values are stated in Table 3 together with the mean Ap index of the respective years. As the correlation coefficient is insensitive to the offsets between two data sets and merely shows how simultaneously they evolve in time, the general flux level offset between modeled and LC data is not contained in the correlation. The correlation is overall good increasing from 0.72 in 2003 to 0.80 in 2008 at the 5.25 L shell. Although the model generally captures short-term variability to a satisfactory degree, it yields better performances in doing so during solar minimum compared to solar maximum for all investigated shells and energy channels.

### 3.4. Maximum Flux Level and Implication for Modeling Multiple Solar Cycles

One of the advantages of the Ap model is that it provides electron flux estimates beyond the limited satellite measurements as long as the Ap index is available from 1932, or a reconstructed version that goes back to ~1850 (Matthes et al., 2017). The model itself, however, is only based on parts of solar cycles 23 and 24. As these are relatively weak cycles compared to previous solar cycles, the evident flux saturation at times of high Ap index might impact the model's performance representing multiple solar cycles.

Figure 5a shows how the modeled >43 keV flux depends on the Ap index on different L shells. It is evident that fluxes reach a saturation level for high Ap values, meaning that their sensitivity to variations in Ap is diminished. As can be seen, fluxes on L shell 5.25 barely respond to changes in Ap which happen above an Ap value of approximately 40 explaining the plateau features seen in Figure 3a. The electron flux on lower L shells (4.25) saturates at Ap levels around 60. For high L shells (7.75) there is a weak flux responses for Ap larger than 5. Maximum flux level for the >43 keV electrons is approximately $2 \cdot 10^5$ cm$^{-2}$·s$^{-1}$·sr$^{-1}$ for L shells 4.25 and 5.25, while it is approximately $1 \cdot 10^5$ cm$^{-2}$·s$^{-1}$·sr$^{-1}$ for L shell 7.75. These features are different in the LC fluxes >43 keV shown in Figure 1. For example, for high Ap values >40 the flux levels are not systematically different for L shells between 4 and 8. The flux levels do not reach a plateau but increase about an order of magnitude between Ap values 30 and 80. (For Ap > 80 the statistics is poor, and it is hard to evaluate the Ap dependence.)

Figure 5b shows how the modeled >114-keV flux depends on the Ap index for different L values. At L shell 5.25, the fluxes increase with increasing

**Table 3**
*Correlation Coefficients Between Modeled and MEPED LC Fluxes for Different Years, L Shells, and Energy Channels*

| Year/ L shell | Correlation > 43 keV LC flux/Ap model | | | Correlation >114 keV LC flux/Ap model | | | Mean Ap |
|---|---|---|---|---|---|---|---|
| | 4.25 | 5.25 | 7.75 | 4.25 | 5.25 | 7.75 | |
| 2003 | 0.65 | 0.72 | 0.47 | 0.74 | 0.49 | 0.43 | 22 |
| 2005 | 0.77 | 0.75 | 0.49 | 0.77 | 0.65 | 0.41 | 13 |
| 2008 | 0.80 | 0.80 | 0.72 | 0.82 | 0.76 | 0.52 | 7 |

*Note.* The mean Ap index for the different years is listed.

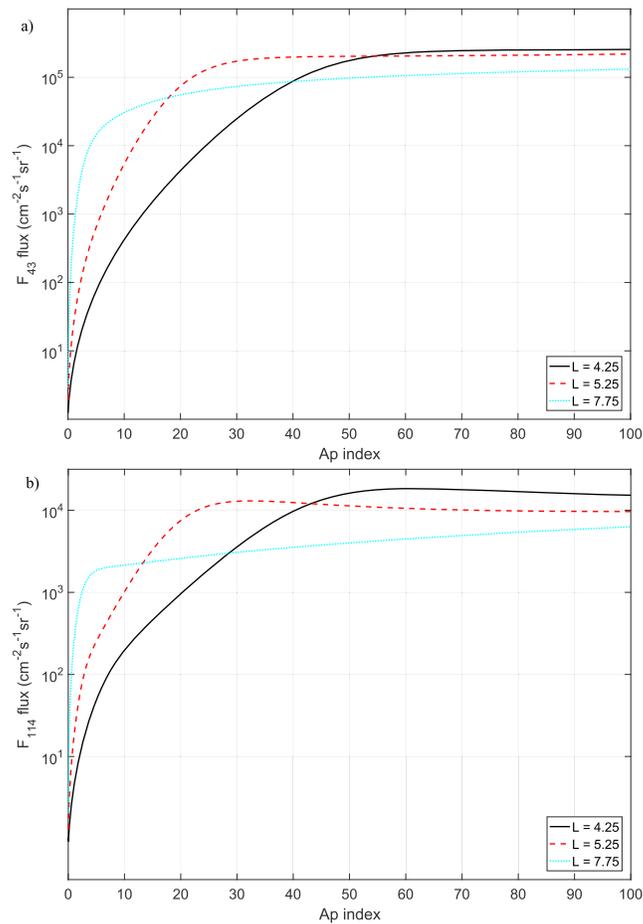

**Figure 5.** Illustration of the Ap sensitivity of the flux levels themselves for (a) the >43-keV fluxes and (b) the >114-keV fluxes.

Ap until the index reaches a value of approximately 30. In this case, however, fluxes do not just saturate and reach a limiting flux level but show a decrease after passing an Ap value of 30. Again, for L shell 7.75, there is a weak Ap dependence associated with Ap values larger than 5. Maximum flux levels >43-keV electrons are approximately $2 \cdot 10^4$ cm$^{-2}\cdot$s$^{-1}\cdot$sr$^{-1}$ for L shells 4.25 and 5.25, while it is approximately $7 \cdot 10^3$ cm$^{-2}\cdot$s$^{-1}\cdot$sr$^{-1}$ for L shell 7.75. The LC fluxes in Figure 1 do not show a strong saturation effect, and there is no evidence for a local maxima.

Figure 6 shows the number of days in each year from 1970 to 2016 with a daily Ap-index >40. These would be the days affected by dampened model sensitivity. The period from 2002 to 2012 which forms the data base of the Ap model is marked by the grey area. There are almost 100 days throughout the whole period 2002–2012 associated with Ap values >40. Still, previous solar cycles exhibit a larger amount of these days and will thus be more strongly affected by saturation effects.

## 4. Discussion

The Ap model by van de Kamp et al. (2016) aims to enable simulations of EEP impacts on the atmosphere with realistic MEE variability. Based on the comparison between the Ap model fluxes and the LC fluxes estimated from measured particle fluxes, we assess the general flux level, evaluate the model performance in terms of the variability within a solar cycle and its capability of representing the variability of previous solar cycles.

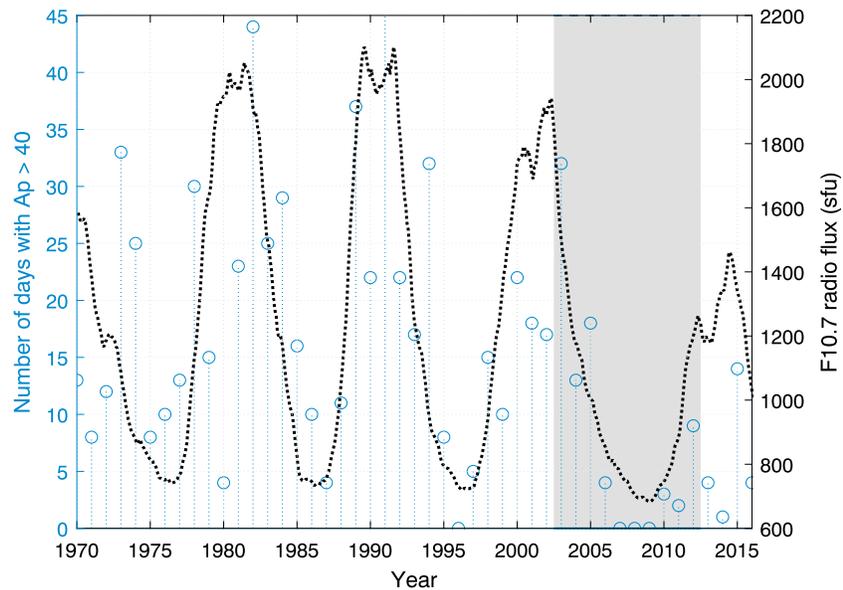

**Figure 6.** Number of days during each year with a daily Ap index >40 from 1970 to 2016. The black dotted line indicates the solar cycle given by the F10.7 radio flux and the grey area marks years considered in the Ap model derivation.

Comparison with the estimated LC fluxes shows a general underestimation of MEE fluxes in the Ap model. The level of underestimation varies with the flux level as shown in Figure 1 and type of storm, L shell, and solar cycle as shown in Figure 3. Evaluating the flux variability within a solar cycle reveals that the Ap model exaggerates the relative flux level difference between solar maximum and solar minimum as quantified in Table 2. Tu et al. (2010) found a short electron lifetime in the radiaton belt due to fast precipitation in the order of hours (strong electron diffusion) at all energies for event studies including weak, moderate, and strong geomagnetic storms. Shprits et al. (2005) give the electron lifetime in the outer radiation belt as a function of the Kp index (3/Kp). Shorter lifetime is associated with stronger level of pitch angle diffusion, influencing the level of pitch angle anisotropy. Hence, with fewer storms and lower mean Ap values as shown in Table 2, there might be a systematic difference in the level of pitch angle anisotropy throughout the solar cycle, but a firm conclusion needs further investigation. Fitting a model to the LC fluxes instead of the 0° fluxes will solve the uncertainty related to strong pitch angle anisotropy. The day to day flux variability in Figures 3 and 4 revealed, however, that the model appears inadequate to represent the recovery periods of CIR storms possibly causing a systematic bias in the solar cycle variability.

To evaluate the length of the precipitation events, a short evaluation routine is implemented which sorts out precipitation events based on variations of the Ap index and examines the corresponding event length of the elevated fluxes. Only events which were preceded by a 2-day-long quiet period were considered. The start and end of a storm were defined by the times fluxes dropped below values 1 order of magnitude lower than the event peak fluxes (after removing flux base lines). This routine works well in terms of picking out and examining single events in 2005 and 2008. The year 2003 was, however, marked by very elevated geomagnetic activity, eliminating most events as they did not exhibit a quiet period in advance. In 2008 a total of 18 isolated events were identified including the event depicted in Figure 4. The LC fluxes imply a 4-day longer event length than the event length of the Ap-modeled fluxes in this case. As this corresponds to a 114% longer period of elevated flux levels, significant differences in the overall energy input by MEE are to be expected. Indeed, LC fluxes suggest a 50% longer event time when averaging all 18 identified events in 2008. For the L shells 5–5.5, 11 events could be identified in 2005, yielding on average a 34% longer event time for LC fluxes than for modeled fluxes. This behavior is especially intriguing as this underestimation of MEE impact arises during the declining and minimum phase in the solar cycle. This is consistent with the relative larger differences found between the LC and Ap-modeled fluxes in Table 2.

The Ap index shows relatively low values in the recovery phase of the CIR storms as illustrated in Figure 4 and this was also shown in the superposed epoch analysis by Ødegaard et al. (2017). The Ap index is derived from measurements of magnetic deviations at midlatitudes caused by currents flowing in the ionosphere/magnetosphere system. Precipitating particles depositing their energy below 90 km may not predominantly affect currents due to high resistivity at these heights, nor is there evidence for presenting a direct link between geomagnetic activities and wave-particle interaction that can cause energy and pitch angle diffusion. In fact, MEE precipitation tends to occur in association with the diffuse/pulsating aurora, but the magnetic field deviations are moderately small at the time as seen in the substorm recovery phase (e.g., Partamies et al., 2017). Hence, any models based on the Ap index are bound to cause some ambiguities when assessing MEE. This implies that solving this systematic bias in the Ap model calls for a parameterization with an accumulated Ap value or where the Ap dependence is set in a context. It is not only a specific Ap value associated to a specific flux level, but the context the Ap value occurs in is also important. Hence, it is possible that part of the yearly bias can be solved by changing the model function and its dependency on Ap.

In terms of addressing the model's capability of representing previous solar cycles, Figure 5 shows that there is an upper limit to the daily flux level after a certain Ap value has been reached. Based on the comparison with the LC fluxes, it appears that the upper limit is generally too low and suppresses the variability associated with high Ap values (>40). The empirical flux data on which the model is based, is from year 2002 to 2012. This period has relatively few days that reach the saturation level compared to previous cycles as shown in Figure 6. This raises a question: How will an empirical database from relatively weak geomagnetic activity influence the model's ability to predict flux level and variability of previous solar cycles?

Maliniemi et al. (2013) shows MEE fluxes (30–100 and 100–300 keV) averaged over the three winter months, December, January, and February, in the Northern Hemisphere above 40°N for 1980–2010 covering almost three solar cycles. They use fluxes only from the 0° detector, and their results might therefore be comparable to the Ap model. The electron fluxes show a clear solar cycle variation with the maximum fluxes observed in the declining phases of the solar cycle in 1985, 1994, and 2004. The maximum flux levels for 30–100 keV varies up to ~50% from one solar cycle to another. Compared with the corresponding Ap index they show an offset of 1–2 years between maximum Ap and maximum electron flux at 30–100 keV. The 100–300 keV electron fluxes show a poorer correlation with the Ap index compared to the 30–100 keV electron flux, from which they conclude that the Ap index is a rather crude proxy for the electron fluxes at these energies.

Based on the Ap-driven MEE ionization at ~80 km (corresponding to 30–100 keV fluxes) shown in Figure 14 in Matthes et al. (2017), the highest impact in the period 1980–2010 at L shells 5.25 and 7.25 is found in 1983–1984, 1991–1992, and 2003–2004. In contrast to the variability found by Maliniemi et al. (2013), the maximum yearly averages are quite similar for all solar cycles in the period 1955–2015, which might be due to the saturation effect in the model. The Ap index during the maximum years exceeds 40 for more than 25 days. The largest difference between the different solar cycles is related to solar minimum values where the model has a higher degree of sensitivity. Hence, it should be noted that parameters like flux strength and variability are likely not adequately represented by the Ap model in a long-term perspective.

It is not evident how to solve the problems associated with saturated flux levels in an empirical model. Fitting the model to the LC fluxes would improve the model's performance in regard to increasing the general flux levels, but as long as the empirical flux data are retrieved during weak solar cycles, underestimation will be an issue. One solution could be as in Maliniemi et al. (2013) to use the full NOAA/POES and EUMETSAT/MetOp series. The first two decades, however, where the solar activity is stronger, the magnetic local time coverage of the satellites are poor, which will cause a local time bias in the electron fluxes. On the other hand, it is possible to use the maximum LC fluxes in this period as an upper limit.

Part of the saturation problem might also be linked to methods of averaging. The model aims to enable simulations of EEP impact on the atmosphere and climate. On a physical basis continuing applying the median neglects the nonzero effects of the minority high-flux data. If the aim of the model was to estimate the most commonly appearing fluxes in a certain time bin/L shell, the median would give the best estimate. When, however, analyzing electron fluxes and their effect in the atmosphere over a certain time, it is misleading to apply the median. Very high fluxes will yield strong atmospheric effects, but are completely neglected by the median as long as they do not account for more than 50% of the data. The same argumentation holds true for very low fluxes. With respect to their effect, all fluxes are of equal importance and one should therefore use a statistical method that reflects the whole flux spectrum.

## 5. Summary and Conclusions

Quantitative measurements of the MEE precipitation are an outstanding question and a key to resolve the total EPP impact on the atmosphere. Accurate quantification of the MEE precipitation, however, remains due to instrumental challenges. MEPED consist of two electron and two proton telescopes, pointed in two directions, approximately 0° and 90° to the local vertical. The level of pitch angle anisotropy varies significantly with particle energy, location, and geomagnetic activity. This implies that the 0° and 90° telescopes alone cannot be used to determine the level of precipitating particle fluxes. The 0° detector measurements will underestimate, while the 90° detector measurements will overestimate the flux of precipitating electrons.

Van de Kamp et al. (2016) provides a model for 30–1,000 keV EEP based on the 0° detector in the period 2002–2012 scaled to daily resolution of the geomagnetic index Ap. The model has been recommended as part of the Solar Forcing for CMIP6 (v3.2) (Matthes et al., 2017) and will as such have a potential wide impact in studies estimating the effects of EPP upon the atmosphere. In order to assess the accuracy of the van de Kamp et al. (2016) model, we compare the modeled electron fluxes with the novel LC flux estimates by Nesse Tyssøy et al. (2016). The main purpose of this study is hence to understand the potential uncertainty in the EPP effect applying the CMIP-6 particle energy input in order to assess the subsequent impact on the atmosphere. The result of the comparison can be summarized as follows:

An overall underestimation of basic flux strength about one order of magnitude arises from utilizing 0° detector electron fluxes instead of the LC flux estimates.

The degree to which the model is able to reproduce general flux levels and short-term variabilities is dependent on the chosen phase of the solar cycle. During solar maximum in 2003, the Ap-modeled fluxes (>43 keV) were ~30% of the estimated LC fluxes (general median difference), while in 2005 and 2008 the Ap-modeled flux levels were ~10% of the estimated LC fluxes.

Although the flux levels in all cases are underestimated, the correlations between modeled data sets and LC flux data sets are generally high with correlations of 0.72, 0.75, and 0.80 for L shell 5.25 for 2003, 2005, and 2008 for >43 keV fluxes, respectively.

In general, the correlation is best in the declining and minimum phase of the solar cycle where the general flux discrepancy is largest. This feature reflects the model's enhanced sensitivity to small Ap values. The year 2008 was characterized by a sequence of CIR storms which were associated with relative weak Ap values (<40). The model generally captures the initial phase of the storm fluxes, but falls short in respect to reproducing elevated flux levels during the recovery phase of CIR-driven storms.

The correlation is worse in the maximum phase of the solar cycle where the general flux discrepancy is least. The comparison between the two flux estimates shows that the Ap model fluxes (>43 keV) reach a plateau where the Ap model fails to reproduce further elevated flux levels and flux variability.

Hence, the Ap model fails in general to reproduce the flux level variability associated with the strongest CME storms (Ap > 40), characterizing solar maximum, and the duration of the CIR storms causing a systematic bias within a solar cycle.

As the Ap model fluxes reach a plateau for Ap > 40, the model's ability to reflect the flux level of previous solar cycles, which were associated with generally higher Ap values throughout the entire cycle, might be questioned.

Given that the Ap-based parametrization is the first model to attempt ascribing electron fluxes to Ap values, its overall performance can be considered successful. Its major advantage compared to satellite measurements is its very long time span, as the Ap index can be reconstructed until year 1850 (Matthes et al., 2017), whereas satellite measurements cover merely the past four decades. The Ap model shows great potential with regard to improvement of temporal variability if some of its cavities in regard to the general flux level, ability to reproduce strong CME, and weak CIR storms are addressed. This can be achieved by fitting the model function to estimated LC fluxes and not against the flux measurement by the 0° detector. Further, the model function should also be evaluated concerning the asymptotic maximum flux level and lack of context in terms of reflecting the CIR recovery phase.

With respect to the current version of the CMIP6 MEE parameterization, the user should be aware that the MEE impact most likely is strongly underestimated in terms of the general flux level. It might exaggerate the difference between the maximum and minimum phase of the solar cycle, but underestimate relative difference between strong and weak solar cycles.

**Acknowledgments**

This study was supported by the Research Council of Norway under contracts 263008 and 223252/F50. The authors thank the NOAA's National Geophysical Data Center (NGDS) for providing NOAA data (http://satdat.ngdc.noaa.gov/) and SPDF Goddard Space Flight Center for geomagnetic Ap indices (http://omniweb.gsfc.nasa.gov/).

Journal of Geophysical Research: Space Physics

10.1029/2018JA025745Rozanov, E., Calisto, M., Egorova, T., Peter, T., & Schmutz, W. (2012). Influence of the precipitating energetic particles on atmospheric chemistry and climate. *Surveys in Geophysics*, *33*(3-4), 483–501. https://doi.org/10.1007/s10712-012-9192-0

Rozanov, E., Callis, L., Schlesinger, M., Yang, F., Andronova, N., & Zubov, V. (2005). Atmospheric response to NOy source due to energetic electron precipitation. *Geophysical Research Letters*, *32*, L14811. https://doi.org/10.1029/2005GL023041

Rusch, D. W., Gérard, J.-C., Solomon, S., Crutzen, P. J., & Reid, G. C. (1981). The effect of particle precipitation events on the neutral and ion chemistry of the middle atmosphere. I - Odd nitrogen. *Planetary and Space Science*, *29*(7), 767–774. https://doi.org/10.1016/0032-0633(81)90048-9

Sandanger, M. I., Ødegaard, L.-K. G., Nesse Tyssøy, H., Stadsnes, J., Søraas, F., Oksavik, K., & Aarsnes, K. (2015). In-flight calibration of NOAA POES proton detectors—Derivation of the MEPED correction factors. *Journal of Geophysical Research: Space Physics*, *120*, 9578–9593. https://doi.org/10.1002/2015JA021388

Seppälä, A., Lu, H., Clilverd, M. A., & Rodger, C. J. (2013). Geomagnetic activity signatures in wintertime stratosphere wind, temperature, and wave response. *Journal of Geophysical Research: Atmospheres*, *118*, 2169–2183. https://doi.org/10.1002/jgrd.50236

Seppälä, A., Randall, C. E., Clilverd, M. A., Rozanov, E., & Rodger, C. J. (2009). Geomagnetic activity and polar surface air temperature variability. *Journal of Geophysical Research*, *114*, A10312. https://doi.org/10.1029/2008JA014029

Shprits, Y. Y., Thorne, R. M., Reeves, G. D., & Friedel, R. (2005). Radial diffusion modeling with empirical lifetimes: Comparison with CRRES observations. *Annales de Geophysique*, *23*(4), 1467–1471. https://doi.org/10.5194/angeo-23-1467-2005

Sinnhuber, M., Nieder, H., & Wieters, N. (2012). Energetic particle precipitation and the chemistry of the mesosphere/lower thermosphere. *Surveys in Geophysics*, *33*(6), 1281–1334. https://doi.org/10.1007/s10712-012-9201-3

Smith-Johnsen, C., Nesse Tyssøy, H., Hendrickx, K., Orsolini, Y., Kishore Kumar, G., Ødegaard, L.-K. G., et al. (2017). Direct and indirect electron precipitation effect on nitric oxide in the polar middle atmosphere, using a full-range energy spectrum. *Journal of Geophysical Research: Space Physics*, *122*, 8679–8693. https://doi.org/10.1002/2017JA024364

Solomon, S., Rusch, D., Gérard, J., Reid, G., & Crutzen, P. (1981). The effect of particle precipitation events on the neutral and ion chemistry of the middle atmosphere: II. Odd hydrogen. *Planetary and Space Science*, *29*(8), 885–893. https://doi.org/10.1016/0032-0633(81)90078-7

Theodoridis, G. C., & Paolini, F. R. (1967). Pitch angle diffusion of relativistic outer belt electrons. *Annales de Geophysique*, *23*, 375–381.

Tsurutani, B. T., Gonzalez, W. D., Gonzalez, A. L. C., Guarnieri, F. L., Gopalswamy, N., Grande, M., Kamide, Y., et al. (2006). Corotating solar wind streams and recurrent geomagnetic activity: A review. *Journal of Geophysical Research*, *111*, A07S01. https://doi.org/10.1029/2005JA011273

Tu, W., Selesnick, R., Li, X., & Looper, M. (2010). Quantification of the precipitation loss of radiation belt electrons observed by SAMPEX. *Journal of Geophysical Research*, *115*, A07210. https://doi.org/10.1029/2009JA014949

van de Kamp, M., Seppälä, A., Clilverd, M. A., Rodger, C. J., Verronen, P. T., & Whittaker, I. C. (2016). A model providing long-term datasets of energetic electron precipitation during geomagnetic storms. *Journal of Geophysical Research: Atmospheres*, *121*, 12,520–12,540. https://doi.org/10.5709/acp-0196-2

Yando, K., Millan, R. M., Green, J. C., & Evans, D. S. (2011). A Monte Carlo simulation of the NOAA POES Medium Energy Proton and Electron Detector instrument. *Journal of Geophysical Research*, *116*, A10231. https://doi.org/10.1029/2011JA016671

Zhang, Y., Sun, W., Feng, X. S., Deehr, C. S., Fry, C. D., & Dryer, M. (2008). Statistical analysis of corotating interaction regions and their geoeffectiveness during solar cycle 23. *Journal of Geophysical Research*, *113*, A08106. https://doi.org/10.1029/2008JA013095
TYSSØY ET AL.

642